\documentclass[%
 reprint,
superscriptaddress,
reprint,
 amsmath,amssymb,
 aps,
 prl,
floatfix,
]{revtex4-2}

\usepackage{graphicx}
\usepackage{dcolumn}
\usepackage{bm}
\usepackage{hyperref}

\def\zrte{ZrTe$_5$}

\usepackage[dvipsnames]{xcolor}
\usepackage{soul}
\usepackage{upgreek}

\begin{document}


\title{Anomalous Hall effect at the Lifshitz transition in ZrTe$_5$}

\author{P.\ M.\ Lozano}\thanks{These authors contributed equally}\affiliation{Department of Physics and Astronomy,
Stony Brook University, Stony Brook, New York 11794-3800, USA}
\affiliation{Condensed Matter Physics and Materials Science Division, 
Brookhaven National Laboratory, Upton, New York 11973-5000, USA}

\author{Gabriel Cardoso}\thanks{These authors contributed equally}\affiliation{Department of Physics and Astronomy,
Stony Brook University, Stony Brook, New York 11794-3800, USA}

\author{Niraj Aryal}
\author{D.\ Nevola}
\author{Genda Gu}
\author{Alexei Tsvelik}
\author{Weiguo Yin}\affiliation{Condensed Matter Physics and Materials Science Division, 
Brookhaven National Laboratory, Upton, New York 11973-5000, USA}
\author{Qiang Li}\email{qiangli@bnl.gov} \affiliation{Department of Physics and Astronomy,
Stony Brook University, Stony Brook, New York 11794-3800, USA}
\affiliation{Condensed Matter Physics and Materials Science Division, 
Brookhaven National Laboratory, Upton, New York 11973-5000, USA}

\date{\today}

\begin{abstract} 
Zirconium pentatelluride \zrte{} is a topological semimetal. The presence of a temperature induced Lifshitz transition, in which the Fermi level goes from the conduction band to the valence band with increasing temperature, provides unique opportunities to study the interplay between Fermi-surface topology, dynamics of Dirac fermions, and Berry curvature in one system. Here we present a combined experimental and theoretical study and show that a low energy model can be used to understand the complicated Hall response and large anomalous Hall effect observed in \zrte{} over a wide range of temperature and magnetic field. We found that the the anomalous Hall contribution dominates the Hall response in a narrow temperature window around the Lifshitz transition, away from which the orbital contribution dominates. Moreover, our results indicate that a  topological phase transition coexists with the Lifshitz transition. Our model provides a unifying framework to understand the Hall effect in semimetals with large Zeeman splitting and non-trivial topology.

\end{abstract}

\maketitle


Zirconium pentatelluride \zrte{} lies at the boundary between a strong and a weak topological insulator, where the gapless state corresponds to a Dirac semimetal \cite{aryal2021topological, zhang2021observation, Weng2014, Chen2017Spectroscopic, mutch2019evidence, Vaswani2020light, Fan2017Transition}.  The non-trivial topology of electronic bands and dynamics of Dirac fermions in \zrte{} have generated considerable attention recently by bringing together chiral magnetic effect \cite{li2016chiral}, three-dimensional quantum Hall effect \cite{tang2019three}, and anomalous Hall effect (AHE) \cite{liang2018anomalous, liu2021induced, mutch2021abrupt} in a single  material. Recent ultrafast terahertz experiments have found giant dissipationless chiral photocurrent arises at the temperature where AHE was detected in the same \zrte{} single crystals \cite{Vaswani2020light, luo2021light}.

\begin{figure*}
    \centering
    \includegraphics{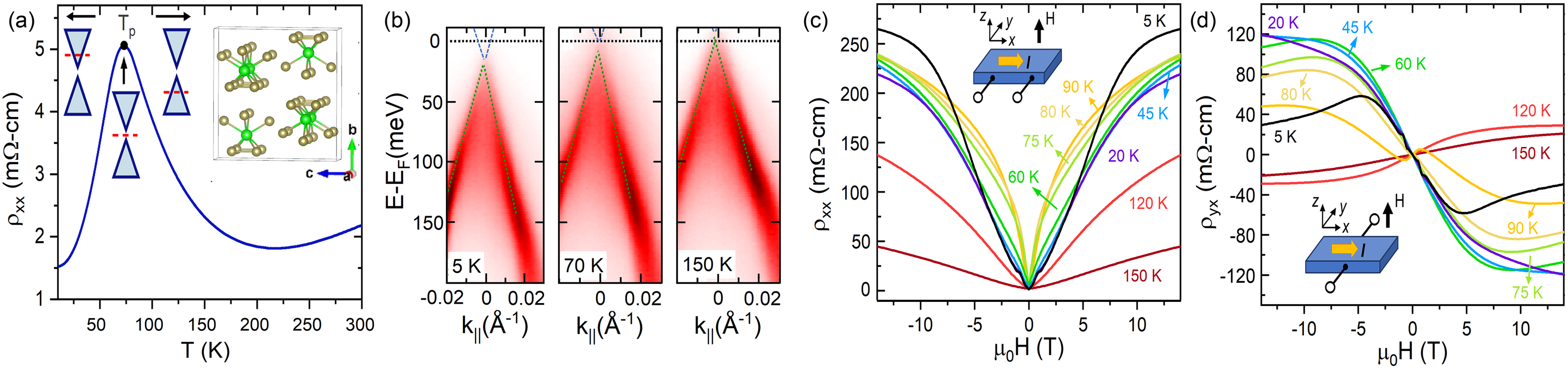}
    \caption{\textbf{(a)} Temperature dependence of the longitudinal resistivity of \zrte{} along the  a-axis. The resistivity peak at $T = T_p$ marks the temperature induced Lifshitz transition. \textbf{(b)} The ARPES of the dispersion near the $\Gamma$ point at 5 K, 70 K and 150 K. The 70K and 150K data were divided by the Fermi-Dirac function to make the data near the Fermi level clearer, the dotted lines are guides for the eye. \textbf{(c,d)} Longitudinal and transverse magnetoresistivity  across the Lifshitz transition. The insets show the measurement configurations.} 
    
    \label{fig1}
\end{figure*}

In non-magnetic materials, such as \zrte{} and $\text{Cd}_3\text{As}_2$, the AHE is generally attributed to the intrinsic Berry curvature of the electronic bands  \cite{nagaosa2010, karplus1954}. In a magnetic field, \zrte{} has a large Zeeman splitting with Land\'{e} g-factor of $\sim$ 21, which in turns produces a  large AHE \cite{liu2016zeeman, choi2020zeeman, sun2020large, liu2021induced}. Another distinctive feature of \zrte{} is its anomalous resistivity peak at temperature $T_p$ ($\approx$ 75 K in our sample), shown in Fig. \ref{fig1}(a).  The resistive anomaly has its origin in the temperature induced Lifshitz transition, in which the Fermi level goes from the conduction to the valence band with increasing temperature \cite{chi2017lifshitz, fu2020dirac, xu2018temperature, okada1980giant, jones1982thermoelectric,Tritt1999large,zhang2017electronic, Zheng2016Transport, pariari2017coexistence, tian2019dirac} and can be understood in the following way. As the Fermi level approaches the small band gap, the density of states is  reduced, and thus suppresses conductivity. This behavior is schematically shown in the left insets of Fig. \ref{fig1}(a), while Fig. \ref{fig1}(b) shows the actual band dispersion taken by angle-resolved photoemission spectroscopy (ARPES) at different temperatures on samples used in the resistivity measurements. The ARPES data were taken along a momentum cut parallel to the conducting chain axis, the crystallographic a-axis, with incident photon energy of 6 eV and an overall resolution of 4meV. \cite{supplementary}.

The large Zeeman splitting, the strong Berry curvature, and the presence of a temperature induced Lifshitz transition produce a complicated temperature and magnetic field dependence of the Hall response in \zrte{}, shown in \ref{fig1}(d), and it is poorly understood. To understand this, we performed a combined experimental and theoretical study to investigate various factors that contribute to the Hall response. Using a low energy model, we quantify the Berry curvature contribution and the orbital contribution to the Hall response as a function of magnetic field at different temperatures. We identified that the Berry curvature makes the dominant contribution to the Hall response in the vicinity of the Lifshitz transition, where the orbital contribution is significantly smaller due to the reduced density of states. Far away from the Lifshitz transition, the orbital contribution  dominates the Hall response, while the contribution from the Berry curvature induced AHE is subdued. This low energy analytical model reconciles various observations in \zrte{}.

Experimentally, the transport properties of \zrte{} were investigated with the current applied along the a-axis of the single crystals and the magnetic field applied parallel to the b-axis, which is the $z$ axis in the insets of Fig.\ref{fig1}(b,d)) \cite{supplementary}. Both the longitudinal resistivity $\rho_{xx}$ and Hall resistivities $\rho_{yx}$ are measured as a function of magnetic field at various temperatures, and they are shown in Fig.\ref{fig1} (c,d). In the high field regime, the Hall resistivity becomes non-linear, signaling the presence of the AHE. This AHE was observed at temperatures as high as 190K, which coincides with the onset temperature the light induced chiral current was observed in the same samples \cite{luo2021light}. The non-linear contribution to the Hall response gets stronger as temperature decreases. Of note, there is a sign change in the Hall coefficient as the temperature goes through the Lifshitz transition, corresponding the dominant charge carriers type changing from electron-like to hole-like.

To understand the complicated temperature and magnetic field dependence of Hall response in \zrte{}, we investigate the Hall conductivity, which can be directly compared with theoretical predictions. The Hall conductivity can be calculated from the magnetoresistivity measurements via the relation $\sigma^{xy}=\rho_{yx}/(\rho_{xx}^2+\rho_{yx}^2)$ \footnote{Here, it is assumed that $\rho_{xx}\sim\rho_{yy}$, as shown in a number of recent works \cite{liu2021induced, mutch2021abrupt}}. The total measured Hall conductivity consists of two contributions, the orbital one and the anomalous one, $\sigma^{xy}=\sigma_{\text{orbital}}^{xy}+\sigma_{\text{AHE}}$ \cite{nagaosa2010}, where the orbital contribution $\sigma_{\text{orbital}}^{xy}$ comes from the Drude formula \cite{pippard1989magnetoresistance}
\begin{equation}
    \sigma^{xy}_{\text{orbital}}=\frac{en\upmu^2 B}{1+\upmu^2B^2}.
    \label{eq:drude}
\end{equation}

\begin{figure*}
    \centering
    \includegraphics{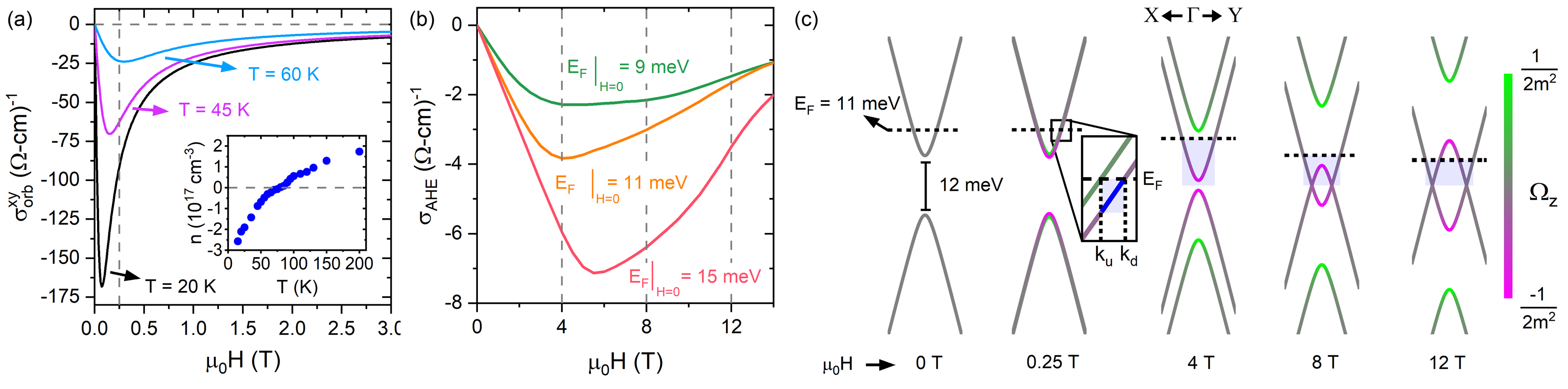}
    \caption{ \textbf{(a,b)}  Magnetic field dependence of the calculated $\sigma_{\text{orbital}}^{xy}$  below $T_p$ for several temperatures and the calculated $\sigma_{\text{AHE}}$ for different zero field Fermi levels of \zrte{}, respectively. The charge carrier density $n$ shown in the inset plot of (a). \textbf{(c)} The calculated electronic band structure of \zrte{} near the $\Gamma$ point, together with the calculated $z$ component of the Berry curvature $\Omega_z$, (values shown in the color scale), the  dotted line represents the position of the Fermi level.} 
    \label{fig2}
\end{figure*}

In the low field limit, $\sigma_{\text{orbital}}^{xy}\sim{B}$, while in the high field limit, $\sigma_{\text{orbital}}^{xy}\sim 1/B$. The maximum value of $\sigma_{\text{orbital}}^{xy}$ occurs at $B=\upmu^{-1}$. Given the high mobility $\upmu \sim 10^5\text{ cm}^2/\text{Vs}$ in \zrte{}, the maximum $\sigma_{\text{orbital}}^{xy}$ appears at $B\sim$ 0.2 T. The overall magnitude of $\sigma_{\text{orbital}}^{xy}$ depends on the carrier density $n$. Fig. \ref{fig2}(a) shows the magnetic field dependence of the calculated $\sigma_{\text{orbital}}^{xy}$ in \zrte{} at three different temperatures.

Now, we turn to the anomalous contribution to the total Hall conductivity, $\sigma^{xy}$, which is given by the sum of the Berry curvature over the occupied states,
\begin{equation}\label{eq:sigmaxy}
    \sigma_{\mathrm{AHE}}=\frac{e^2}{\hbar}\sum_n\int\frac{d^3k}{(2\pi)^3}f(E^n,\mu,T)\Omega_z^n,
\end{equation}
where $\Omega_z^n$ is the $z$ component of the Berry curvature of band $n$, $f$ is the Fermi-Dirac distribution and $\mu$ is the chemical potential. Since the Berry curvature is concentrated at the $\Gamma$ point, which the Fermi level lies close to, it is natural to start with a linearized low energy model of the Hamiltonian. We do this by using the $k \cdot p$ model of \citet{chen2015magnetoinfrared}, consistent with the symmetries of \zrte{},
\begin{equation}
    H_0=m\tau^z+\hbar (v_xk_x\tau^x\sigma^z+ v_yk_y\tau^y+ v_zk_z\tau^x\sigma^x),
\end{equation}
where $m$ is the effective mass of the Dirac fermions, $v_i$ and $k_i$ are the $i$-th components of the Fermi velocity and crystal momentum, respectively, and $\sigma$ and $\tau$ represent the spin and orbital degrees of freedom, respectively. Given the large Zeeman splitting in \zrte{}, the main effect of an external magnetic field can be incorporated by including the Zeeman energy term $H_{Z}=g\mu_B B\sigma^z/2$ in the total Hamiltonian. In this linear approximation, the energies and Berry curvatures of the four bands are given by
\begin{align}
    E^{s_1 s_2}&=s_1\sqrt{(b+s_2\sqrt{m^2+k_{\perp}^2})^2+k^2}\label{eq:eigenvaluesband},\\ 
    \Omega^{s_1 s_2}_z&=-s_2\frac{m}{2(m^2+k_{\perp}^2)^{3/2}},
    \label{eq:omegakdotp}
\end{align}
where the $s_i$ are $\pm$ signs labeling the four bands and the parametrization in energy units $b = g\mu_B B/2$, $k = \hbar v_zk_z$ and $k_{\perp} e^{i\phi} = \hbar(v_xk_x+i v_yk_y)$ was adopted. The details of this calculation can be found in \cite{supplementary}. Eq.(\ref{eq:omegakdotp}) is well-known for the case of a band crossing in 2D systems. Here it shows that the Berry curvature of the four bands in \zrte{} is constant along $k_z$. On a given $k_z$ slice, there is a Chern number-like relation
\begin{equation}
    \int k_{\perp} dk_{\perp} d\phi \,\,\Omega^{s_1 s_2}_z = -\pi s_2.
\end{equation}
This Berry curvature is concentrated on a region of size $m$ around $k_{\perp} = 0$. In the limit $m\to 0$, $\Omega^{s_1 s_2}_z \to -\pi s_2\delta(k_{\perp} e^{i\phi})$. Fig. \ref{fig2}(c) shows the calculated band structure and the value of the $z$ component of the Berry curvature at various magnetic fields. 

The magnetic field dependence of the anomalous Hall contribution from Eq. (\ref{eq:sigmaxy}) can be understood as follows. At zero field, the valence and the conduction band are each doubly degenerate. The Berry curvature of the degenerate bands have equal magnitude but opposite sign that cancel out each other completely, as shown on the left of Fig. \ref{fig2}(c). This leads to zero Berry curvature everywhere in the Brillouin zone, resulting in no AHE. When a magnetic field is applied, the degeneracy is lifted leading to a non-zero Hall contribution from the Berry curvature. The anomalous Hall conductivity is then given by

\begin{align}
    \sigma_{\mathrm{AHE}}=\frac{e^2}{\hbar}\int \dfrac{dk_z dk_{\perp}}{(2\pi)^3} & (f^{++}-f^{+-}+f^{-+}-f^{--}) \nonumber \\ &\times \frac{-mk_{\perp}}{2(m^2+k_{\perp}^2)^{3/2}},
    \label{eq:aheberry}
\end{align}

where we used the short-hand notation $f^{s_1s_2}=f(E^{s_1s_2},\mu,T)$, and the magnetic field dependent chemical potential is determined by fixing the charge carrier density in the system. 

For simplicity, we first consider the zero-temperature limit. In a small field (e.g.\ 0.25 T), the Zeeman energy splits  both the conduction and valence bands into two. The split conduction bands intersect with the Fermi level $E_\text{F}$ at ${k_u}$ and ${k_d}$ for the up-shifted band and down-shifted band, respectively, shown in the inset to Fig. \ref{fig2}(c). At the Dirac point, the Berry curvature makes effectively zero contribution to $\sigma_{\text{AHE}}$ due to the opposite sign of the split conduction bands in this region. This is true for the states with $ k_{\perp} \leq k_u$ on the conduction bands, and is true for all the states on the valence bands. Only the states with $ k_u < k_{\perp} \leq k_d$ (marked by blue in the region) effectively contribute to $\sigma_{\text{AHE}}$. As the field increases, the band splitting increases leading to a smaller value of $k_u$ and a larger value of $k_d$, that gives rise to a rapid increase in $\sigma_{\text{AHE}}$. $k_u$ vanishes when the up-shifted conduction band is above $E_\text{F}$ at ${k_u}$ (e.g.\ at 4 T), at which $\sigma_{\text{AHE}}$ reaches its maximum value. This picture is consistent with the phenomenological use of the hyperbolic tangent to describe the AHE in \zrte{} as done in recent studies \cite{liu2021induced, mutch2021abrupt} and is explained in more detail in \cite{supplementary}. At a higher field (e.g.\ 8 T), at which the Zeeman energy becomes larger than the band gap, a band crossing appears, creating a nodal line. For even higher fields (e.g.\ 12 T), the states with a larger Berry curvature on the up-shifted valence band edge, which is above $E_\text{F}$ at ${k_u}$ now, stop contributing to the AHE and therefore only the states close to the nodal line contribute. This leads to an accelerated decrease in $\sigma_{\text{AHE}}$ with increasing magnetic field. The shaded regions in Fig. \ref{fig2}(c) represent the states that make a net contribution to the AHE. Fig. \ref{fig2}(b) shows the magnetic field dependence of the calculated $\sigma_{\text{AHE}}$ for \zrte{} using Eq. (\ref{eq:aheberry}) with different $E_\text{F}$ values. The higher the Fermi level is in the conduction band, the higher the magnitude of $\sigma_{\text{AHE}}$, due to higher number of participating states. The peak value of $\sigma_{\text{AHE}}$ moves to a higher magnetic field, because a higher Zeeman energy is needed to drive the up-shifted conduction band edge out of $E_\text{F}$.  It is noted that at high magnetic fields, Landau Level quantization is expected in \zrte{}  \cite{chen2015magnetoinfrared, Chen2017Spectroscopic, li2016chiral}. The effect of the Landau level quantization on the results presented herein is discussed in \cite{supplementary}.

\begin{figure*}
    \centering
    \includegraphics{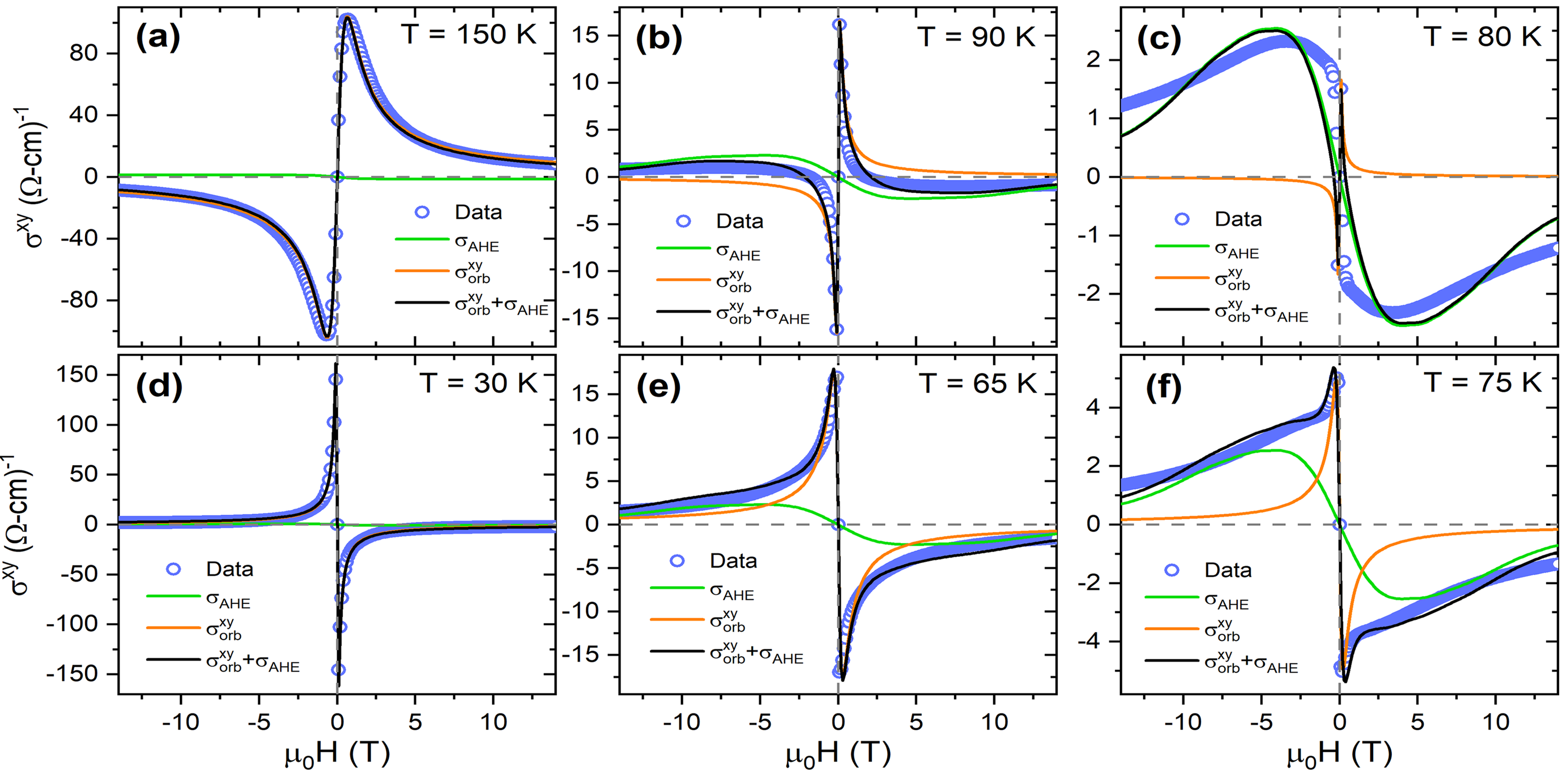}
    \caption{The experimental data (symbols) and the model calculations (lines) of the magnetic field dependence of the Hall conductivity at various temperatures across the Lifshitz transition. \textbf{(a, d)} Far from the transition the Hall conductivity is dominated by the $\sigma_{\text{orbital}}^{xy}$ (represented by the orange lines), overlapped with the total Hall conductivity (black lines). \textbf{(b, e)} As temperature moves closer to $T_p$, the relative contribution of the AHE (green lines) increases, while the $\sigma_{\text{orbital}}^{xy}$ decreases. \textbf{(c, f)} In close proximity to the transition, the $\sigma_{\text{orbital}}^{xy}$ becomes very small in high fields, while the $\sigma_{\text{AHE}}$ dominates the Hall conductivity. The calculations are performed with the 3 meV Fermi level above or below the gap.}
    \label{fig3}
\end{figure*}

Fig. \ref{fig3} shows the experimental results (symbols) and the model calculations (lines) of the magnetic field dependence of the Hall conductivity at various temperatures across the Lifshitz transition. The blue and orange lines are the orbital and the AHE contributions, respectively, to the total Hall conductivity (black lines). For $T \gg T_p$ or $T \ll T_p$, the total Hall conductivity $\sigma^{\text{xy}}$ is dominated by $\sigma_{\text{orbital}}^{xy}$ derived from the classical Drude formula, and $\sigma_{\text{AHE}}$ is negligibly small, relatively speaking. This is shown in Fig. \ref{fig3}(a,d). As temperature approaches $T_p$, $E_\text{F}$ approaches the band edge. $\sigma_{\text{orbital}}^{xy}$ becomes smaller due to reduced carrier density, while $\sigma_{\text{AHE}}$ grows due to concentrated Berry curvature, as shown in Fig. \ref{fig3}(b, e). This is more clearly seen in Fig. \ref{fig3}(b), where the Drude formula (\ref{eq:drude}) cannot explain the sign change $\sigma^{xy}$ at $B \approx \pm 3$ T, and the AHE contribution becomes clear.

It is quite remarkable that the dominance of the AHE in the total Hall response occurs in a narrow temperature window of $\pm$ 2 K around $T_p$. Fig. \ref{fig3}(c, f) show the Hall conductivity for T = 80 K and T = 75 K, respectively. At $T_p \approx 75$ K, the charge carriers of the system are predominantly electrons, as shown in the sign of the orbital contribution, $\sigma_{\text{AHE}}$ is several times more than $\sigma_{\text{orbital}}^{xy}$ at high fields. At 80 K, the switch-over of the charge carriers types occurs, which can be seen from the positive values of the orbital contribution, where the chemical potential is expected to lie closer to the valence band edge.  From 5 K to 200 K, the mobility obtained from fitting the Drude formula is in the order of $10^5\text{ cm}^2/\text{Vs}$, while the carrier density $n$ goes monotonically from $-2.5 \times 10^{17} \text{cm}^{-3}$ to $1.8\times 10^{17} \text{cm}^{-3}$, crossing zero between 75 K and 80 K, in agreement with a previous study on the Lifshitz transition \cite{chi2017lifshitz}. The temperature evolution of the carrier density $n$ is shown in the inset of Fig. \ref{fig2}(a).

We note that the calculated $\sigma_{\text{AHE}}$ in our model uses the chemical potential as the only fitting parameter, while all other parameters use the values reported in the literature: $g = 21.3$ \cite{liu2016zeeman}, the gap is $12$ meV \cite{Chen2017Spectroscopic}, the $x$, $y$ and $z$ Fermi velocities are $9.7 \times 10^5$ m/s, $6.8 \times 10^{5}$ m/s and $9.7 \times 10^4$ m/s \cite{li2016chiral}, respectively, and relevant lattice constants \cite{aryal2021topological}. Near the Lifshitz transition (e.g. $80$ K and $75$ K), we found that the Fermi level positioned at $3$ meV above or below the gap produces a temperature and magnetic field dependence of $\sigma^{\text{xy}}$ that is in good quantitative agreement with the experimental data. The discrepancy between the calculation and the data at very low field (a fraction of 1 T) at $80$ K (Fig. 3c) is also noted. This discrepancy comes from slightly different band parameters above and below $T_p$ that are not accounted for in our model. Affecting the band parameters include band anisotropy, thermal expansion, and particle-hole-asymmetry \cite{aryal2021topological, fu2020dirac}, the mass term \cite{Fan2017Transition, xu2018temperature, zhang2021observation}, and the Fermi velocities \cite{zhang2017electronic}. Nevertheless, over a broad range of temperature and magnetic field, our model can be applied to \zrte{} samples with different $T_p$ and/or gap size, as we show in the supplementary materials \cite{supplementary}. Moreover, our model can be adapted to understand the complicated Hall effect in other Dirac semimetals with large Zeeman splitting.    

Our analysis also shows that, while $\sigma^{xy}_{\text{orb}}$ changes sign at $T_p$, $\sigma_{\text{AHE}}$ does not. The application of the $k \cdot p$ model then requires the use of different signs for the mass term, i.e. $m<0$ for $T<T_p$ and $m>0$ for $T>T_p$. This sign change  shows that the Lifshitz transition involves more than a simple renormalization of the Fermi level. It means that the system goes through a temperature driven topological phase transition between weak and strong topological insulator as suggested in previous studies \cite{Fan2017Transition, xu2018temperature}. In particular, the sign change of the mass term can be induced by many modes of small atomic displacement that preserve the crystal symmetry \cite{aryal2021topological}. We leave this for future studies. 

In summary, we have developed an effective low energy model that treats the classical orbital and the Berry curvature induced anomalous contributions to Hall effects on a equal footing in topological semimetals with a large Land\'{e} g-factor. With the input of the materials' electronic and structural parameters, the model can predict each contribution over a wide range of temperature and magnetic field. It is shown that the model-calculated temperature and magnetic field dependence of the Hall conductivity are in quantitative agreement with the experimental data in \zrte{} having a temperature induced Lifshitz transition. We also discovered that a sign change in the mass term in our calculations is involved as the system goes through the Lifshitz transition. It would be interesting to investigate whether the sign changing is the result of topological phase transition in the electronic structures that coexists with the Lifshitz transition and in particular a subtle structural transition.    

\begin{acknowledgments}
\section{Acknowledgements}
We acknowledge helpful discussions with Alexander Abanov and Jennifer Cano. This work was primarily supported by the U.S. Department of Energy, Office of Basic Energy Sciences, Division of Materials Sciences and Engineering, under Contract No. DE-SC0012704. G.C. was supported by Grant NSF DMR-1606591. This research used resources of the Center for Functional Nanomaterials, which is a U.S. DOE Office of Science Facility, at Brookhaven National Laboratory.

\end{acknowledgments}

\bibliography{apssamp}

\begin{thebibliography}{35}%
\makeatletter
\providecommand \@ifxundefined [1]{%
 \@ifx{#1\undefined}
}%
\providecommand \@ifnum [1]{%
 \ifnum #1\expandafter \@firstoftwo
 \else \expandafter \@secondoftwo
 \fi
}%
\providecommand \@ifx [1]{%
 \ifx #1\expandafter \@firstoftwo
 \else \expandafter \@secondoftwo
 \fi
}%
\providecommand \natexlab [1]{#1}%
\providecommand \enquote  [1]{``#1''}%
\providecommand \bibnamefont  [1]{#1}%
\providecommand \bibfnamefont [1]{#1}%
\providecommand \citenamefont [1]{#1}%
\providecommand \href@noop [0]{\@secondoftwo}%
\providecommand \href [0]{\begingroup \@sanitize@url \@href}%
\providecommand \@href[1]{\@@startlink{#1}\@@href}%
\providecommand \@@href[1]{\endgroup#1\@@endlink}%
\providecommand \@sanitize@url [0]{\catcode `\\12\catcode `\$12\catcode
  `\&12\catcode `\#12\catcode `\^12\catcode `\_12\catcode `\%12\relax}%
\providecommand \@@startlink[1]{}%
\providecommand \@@endlink[0]{}%
\providecommand \url  [0]{\begingroup\@sanitize@url \@url }%
\providecommand \@url [1]{\endgroup\@href {#1}{\urlprefix }}%
\providecommand \urlprefix  [0]{URL }%
\providecommand \Eprint [0]{\href }%
\providecommand \doibase [0]{https://doi.org/}%
\providecommand \selectlanguage [0]{\@gobble}%
\providecommand \bibinfo  [0]{\@secondoftwo}%
\providecommand \bibfield  [0]{\@secondoftwo}%
\providecommand \translation [1]{[#1]}%
\providecommand \BibitemOpen [0]{}%
\providecommand \bibitemStop [0]{}%
\providecommand \bibitemNoStop [0]{.\EOS\space}%
\providecommand \EOS [0]{\spacefactor3000\relax}%
\providecommand \BibitemShut  [1]{\csname bibitem#1\endcsname}%
\let\auto@bib@innerbib\@empty
\bibitem [{\citenamefont {Aryal}\ \emph {et~al.}(2021)\citenamefont {Aryal},
  \citenamefont {Jin}, \citenamefont {Li}, \citenamefont {Tsvelik},\ and\
  \citenamefont {Yin}}]{aryal2021topological}%
  \BibitemOpen
  \bibfield  {author} {\bibinfo {author} {\bibfnamefont {N.}~\bibnamefont
  {Aryal}}, \bibinfo {author} {\bibfnamefont {X.}~\bibnamefont {Jin}}, \bibinfo
  {author} {\bibfnamefont {Q.}~\bibnamefont {Li}}, \bibinfo {author}
  {\bibfnamefont {A.~M.}\ \bibnamefont {Tsvelik}},\ and\ \bibinfo {author}
  {\bibfnamefont {W.}~\bibnamefont {Yin}},\ }\bibfield  {title} {\bibinfo
  {title} {Topological phase transition and phonon-space {D}irac topology
  surfaces in $\text{ZrTe}_5$},\ }\href
  {https://journals.aps.org/prl/abstract/10.1103/PhysRevLett.126.016401}
  {\bibfield  {journal} {\bibinfo  {journal} {Phys. Rev. Lett.}\ }\textbf
  {\bibinfo {volume} {126}},\ \bibinfo {pages} {016401} (\bibinfo {year}
  {2021})}\BibitemShut {NoStop}%
\bibitem [{\citenamefont {Zhang}\ \emph {et~al.}(2021)\citenamefont {Zhang},
  \citenamefont {Noguchi}, \citenamefont {Kuroda}, \citenamefont {Lin},
  \citenamefont {Kawaguchi}, \citenamefont {Yaji}, \citenamefont {Harasawa},
  \citenamefont {Lippmaa}, \citenamefont {Nie}, \citenamefont {Weng} \emph
  {et~al.}}]{zhang2021observation}%
  \BibitemOpen
  \bibfield  {author} {\bibinfo {author} {\bibfnamefont {P.}~\bibnamefont
  {Zhang}}, \bibinfo {author} {\bibfnamefont {R.}~\bibnamefont {Noguchi}},
  \bibinfo {author} {\bibfnamefont {K.}~\bibnamefont {Kuroda}}, \bibinfo
  {author} {\bibfnamefont {C.}~\bibnamefont {Lin}}, \bibinfo {author}
  {\bibfnamefont {K.}~\bibnamefont {Kawaguchi}}, \bibinfo {author}
  {\bibfnamefont {K.}~\bibnamefont {Yaji}}, \bibinfo {author} {\bibfnamefont
  {A.}~\bibnamefont {Harasawa}}, \bibinfo {author} {\bibfnamefont
  {M.}~\bibnamefont {Lippmaa}}, \bibinfo {author} {\bibfnamefont
  {S.}~\bibnamefont {Nie}}, \bibinfo {author} {\bibfnamefont {H.}~\bibnamefont
  {Weng}}, \emph {et~al.},\ }\bibfield  {title} {\bibinfo {title} {Observation
  and control of the weak topological insulator state in $\text{ZrTe}_5$},\
  }\href {https://doi.org/10.1038/s41467-020-20564-8} {\bibfield  {journal}
  {\bibinfo  {journal} {Nat. Commun.}\ }\textbf {\bibinfo {volume} {12}},\
  \bibinfo {pages} {1} (\bibinfo {year} {2021})}\BibitemShut {NoStop}%
\bibitem [{\citenamefont {Weng}\ \emph {et~al.}(2014)\citenamefont {Weng},
  \citenamefont {Dai},\ and\ \citenamefont {Fang}}]{Weng2014}%
  \BibitemOpen
  \bibfield  {author} {\bibinfo {author} {\bibfnamefont {H.}~\bibnamefont
  {Weng}}, \bibinfo {author} {\bibfnamefont {X.}~\bibnamefont {Dai}},\ and\
  \bibinfo {author} {\bibfnamefont {Z.}~\bibnamefont {Fang}},\ }\bibfield
  {title} {\bibinfo {title} {Transition-metal pentatelluride $\text{ZrTe}_5$
  and $\text{HfTe}_5$: A paradigm for large-gap quantum spin {H}all
  insulators},\ }\href {https://doi.org/10.1103/PhysRevX.4.011002} {\bibfield
  {journal} {\bibinfo  {journal} {Phys. Rev. X}\ }\textbf {\bibinfo {volume}
  {4}},\ \bibinfo {pages} {011002} (\bibinfo {year} {2014})}\BibitemShut
  {NoStop}%
\bibitem [{\citenamefont {Chen}\ \emph {et~al.}(2017)\citenamefont {Chen},
  \citenamefont {Chen}, \citenamefont {Zhong}, \citenamefont {Schneeloch},
  \citenamefont {Zhang}, \citenamefont {Huang}, \citenamefont {Qu},
  \citenamefont {Yu}, \citenamefont {Li}, \citenamefont {Gu},\ and\
  \citenamefont {Wang}}]{Chen2017Spectroscopic}%
  \BibitemOpen
  \bibfield  {author} {\bibinfo {author} {\bibfnamefont {Z.~G.}\ \bibnamefont
  {Chen}}, \bibinfo {author} {\bibfnamefont {R.~Y.}\ \bibnamefont {Chen}},
  \bibinfo {author} {\bibfnamefont {R.~D.}\ \bibnamefont {Zhong}}, \bibinfo
  {author} {\bibfnamefont {J.}~\bibnamefont {Schneeloch}}, \bibinfo {author}
  {\bibfnamefont {C.}~\bibnamefont {Zhang}}, \bibinfo {author} {\bibfnamefont
  {Y.}~\bibnamefont {Huang}}, \bibinfo {author} {\bibfnamefont
  {F.}~\bibnamefont {Qu}}, \bibinfo {author} {\bibfnamefont {R.}~\bibnamefont
  {Yu}}, \bibinfo {author} {\bibfnamefont {Q.}~\bibnamefont {Li}}, \bibinfo
  {author} {\bibfnamefont {G.~D.}\ \bibnamefont {Gu}},\ and\ \bibinfo {author}
  {\bibfnamefont {N.~L.}\ \bibnamefont {Wang}},\ }\bibfield  {title} {\bibinfo
  {title} {Spectroscopic evidence for bulk-band inversion and three-dimensional
  massive {D}irac fermions in $\text{ZrTe}_5$},\ }\href
  {https://doi.org/10.1073/pnas.1613110114} {\bibfield  {journal} {\bibinfo
  {journal} {Proc. Nat. Acad. Sci. USA}\ }\textbf {\bibinfo {volume} {114}},\
  \bibinfo {pages} {816} (\bibinfo {year} {2017})}\BibitemShut {NoStop}%
\bibitem [{\citenamefont {Mutch}\ \emph {et~al.}(2019)\citenamefont {Mutch},
  \citenamefont {Chen}, \citenamefont {Went}, \citenamefont {Qian},
  \citenamefont {Wilson}, \citenamefont {Andreev}, \citenamefont {Chen},\ and\
  \citenamefont {Chu}}]{mutch2019evidence}%
  \BibitemOpen
  \bibfield  {author} {\bibinfo {author} {\bibfnamefont {J.}~\bibnamefont
  {Mutch}}, \bibinfo {author} {\bibfnamefont {W.-C.}\ \bibnamefont {Chen}},
  \bibinfo {author} {\bibfnamefont {P.}~\bibnamefont {Went}}, \bibinfo {author}
  {\bibfnamefont {T.}~\bibnamefont {Qian}}, \bibinfo {author} {\bibfnamefont
  {I.~Z.}\ \bibnamefont {Wilson}}, \bibinfo {author} {\bibfnamefont
  {A.}~\bibnamefont {Andreev}}, \bibinfo {author} {\bibfnamefont {C.-C.}\
  \bibnamefont {Chen}},\ and\ \bibinfo {author} {\bibfnamefont {J.-H.}\
  \bibnamefont {Chu}},\ }\bibfield  {title} {\bibinfo {title} {Evidence for a
  strain-tuned topological phase transition in $\text{ZrTe}_5$},\ }\href
  {https://doi.org/10.1126/sciadv.aav9771} {\bibfield  {journal} {\bibinfo
  {journal} {Sci. Adv.}\ }\textbf {\bibinfo {volume} {5}},\ \bibinfo {pages}
  {eaav9771} (\bibinfo {year} {2019})}\BibitemShut {NoStop}%
\bibitem [{\citenamefont {Vaswani}\ \emph {et~al.}(2020)\citenamefont
  {Vaswani}, \citenamefont {Wang}, \citenamefont {Mudiyanselage}, \citenamefont
  {Li}, \citenamefont {Lozano}, \citenamefont {Gu}, \citenamefont {Cheng},
  \citenamefont {Song}, \citenamefont {Luo}, \citenamefont {Kim}, \citenamefont
  {Huang}, \citenamefont {Liu}, \citenamefont {Mootz}, \citenamefont {Perakis},
  \citenamefont {Yao}, \citenamefont {Ho},\ and\ \citenamefont
  {Wang}}]{Vaswani2020light}%
  \BibitemOpen
  \bibfield  {author} {\bibinfo {author} {\bibfnamefont {C.}~\bibnamefont
  {Vaswani}}, \bibinfo {author} {\bibfnamefont {L.~L.}\ \bibnamefont {Wang}},
  \bibinfo {author} {\bibfnamefont {D.~H.}\ \bibnamefont {Mudiyanselage}},
  \bibinfo {author} {\bibfnamefont {Q.}~\bibnamefont {Li}}, \bibinfo {author}
  {\bibfnamefont {P.~M.}\ \bibnamefont {Lozano}}, \bibinfo {author}
  {\bibfnamefont {G.~D.}\ \bibnamefont {Gu}}, \bibinfo {author} {\bibfnamefont
  {D.}~\bibnamefont {Cheng}}, \bibinfo {author} {\bibfnamefont
  {B.}~\bibnamefont {Song}}, \bibinfo {author} {\bibfnamefont {L.}~\bibnamefont
  {Luo}}, \bibinfo {author} {\bibfnamefont {R.~H.~J.}\ \bibnamefont {Kim}},
  \bibinfo {author} {\bibfnamefont {C.}~\bibnamefont {Huang}}, \bibinfo
  {author} {\bibfnamefont {Z.}~\bibnamefont {Liu}}, \bibinfo {author}
  {\bibfnamefont {M.}~\bibnamefont {Mootz}}, \bibinfo {author} {\bibfnamefont
  {I.~E.}\ \bibnamefont {Perakis}}, \bibinfo {author} {\bibfnamefont
  {Y.}~\bibnamefont {Yao}}, \bibinfo {author} {\bibfnamefont {K.~M.}\
  \bibnamefont {Ho}},\ and\ \bibinfo {author} {\bibfnamefont {J.}~\bibnamefont
  {Wang}},\ }\bibfield  {title} {\bibinfo {title} {Light-driven {R}aman
  coherence as a nonthermal route to ultrafast topology switching in a {D}irac
  semimetal},\ }\href {https://doi.org/10.1103/PhysRevX.10.021013} {\bibfield
  {journal} {\bibinfo  {journal} {Phys. Rev. X}\ }\textbf {\bibinfo {volume}
  {10}},\ \bibinfo {pages} {021013} (\bibinfo {year} {2020})}\BibitemShut
  {NoStop}%
\bibitem [{\citenamefont {Fan}\ \emph {et~al.}(2017)\citenamefont {Fan},
  \citenamefont {Liang}, \citenamefont {Chen}, \citenamefont {Yao},\ and\
  \citenamefont {Zhou}}]{Fan2017Transition}%
  \BibitemOpen
  \bibfield  {author} {\bibinfo {author} {\bibfnamefont {Z.}~\bibnamefont
  {Fan}}, \bibinfo {author} {\bibfnamefont {Q.~F.}\ \bibnamefont {Liang}},
  \bibinfo {author} {\bibfnamefont {Y.~B.}\ \bibnamefont {Chen}}, \bibinfo
  {author} {\bibfnamefont {S.~H.}\ \bibnamefont {Yao}},\ and\ \bibinfo {author}
  {\bibfnamefont {J.}~\bibnamefont {Zhou}},\ }\bibfield  {title} {\bibinfo
  {title} {Transition between strong and weak topological insulator in
  $\text{ZrTe}_5$ and $\text{HfTe}_5$},\ }\href
  {https://doi.org/10.1038/srep45667} {\bibfield  {journal} {\bibinfo
  {journal} {Sci. Rep.}\ }\textbf {\bibinfo {volume} {7}},\ \bibinfo {pages}
  {45667} (\bibinfo {year} {2017})}\BibitemShut {NoStop}%
\bibitem [{\citenamefont {Li}\ \emph {et~al.}(2016)\citenamefont {Li},
  \citenamefont {Kharzeev}, \citenamefont {Zhang}, \citenamefont {Huang},
  \citenamefont {Pletikosi{\'c}}, \citenamefont {Fedorov}, \citenamefont
  {Zhong}, \citenamefont {Schneeloch}, \citenamefont {Gu},\ and\ \citenamefont
  {Valla}}]{li2016chiral}%
  \BibitemOpen
  \bibfield  {author} {\bibinfo {author} {\bibfnamefont {Q.}~\bibnamefont
  {Li}}, \bibinfo {author} {\bibfnamefont {D.~E.}\ \bibnamefont {Kharzeev}},
  \bibinfo {author} {\bibfnamefont {C.}~\bibnamefont {Zhang}}, \bibinfo
  {author} {\bibfnamefont {Y.}~\bibnamefont {Huang}}, \bibinfo {author}
  {\bibfnamefont {I.}~\bibnamefont {Pletikosi{\'c}}}, \bibinfo {author}
  {\bibfnamefont {A.}~\bibnamefont {Fedorov}}, \bibinfo {author} {\bibfnamefont
  {R.}~\bibnamefont {Zhong}}, \bibinfo {author} {\bibfnamefont
  {J.}~\bibnamefont {Schneeloch}}, \bibinfo {author} {\bibfnamefont
  {G.}~\bibnamefont {Gu}},\ and\ \bibinfo {author} {\bibfnamefont
  {T.}~\bibnamefont {Valla}},\ }\bibfield  {title} {\bibinfo {title} {Chiral
  magnetic effect in $\text{ZrTe}_5$},\ }\href
  {https://doi.org/10.1038/nphys3648} {\bibfield  {journal} {\bibinfo
  {journal} {Nat. Phys.}\ }\textbf {\bibinfo {volume} {12}},\ \bibinfo {pages}
  {550} (\bibinfo {year} {2016})}\BibitemShut {NoStop}%
\bibitem [{\citenamefont {Tang}\ \emph {et~al.}(2019)\citenamefont {Tang},
  \citenamefont {Ren}, \citenamefont {Wang}, \citenamefont {Zhong},
  \citenamefont {Schneeloch}, \citenamefont {Yang}, \citenamefont {Yang},
  \citenamefont {Lee}, \citenamefont {Gu}, \citenamefont {Qiao},\ and\
  \citenamefont {Zhang}}]{tang2019three}%
  \BibitemOpen
  \bibfield  {author} {\bibinfo {author} {\bibfnamefont {F.}~\bibnamefont
  {Tang}}, \bibinfo {author} {\bibfnamefont {Y.}~\bibnamefont {Ren}}, \bibinfo
  {author} {\bibfnamefont {P.}~\bibnamefont {Wang}}, \bibinfo {author}
  {\bibfnamefont {R.}~\bibnamefont {Zhong}}, \bibinfo {author} {\bibfnamefont
  {J.}~\bibnamefont {Schneeloch}}, \bibinfo {author} {\bibfnamefont {S.~A.}\
  \bibnamefont {Yang}}, \bibinfo {author} {\bibfnamefont {K.}~\bibnamefont
  {Yang}}, \bibinfo {author} {\bibfnamefont {P.~A.}\ \bibnamefont {Lee}},
  \bibinfo {author} {\bibfnamefont {G.}~\bibnamefont {Gu}}, \bibinfo {author}
  {\bibfnamefont {Z.}~\bibnamefont {Qiao}},\ and\ \bibinfo {author}
  {\bibfnamefont {L.}~\bibnamefont {Zhang}},\ }\bibfield  {title} {\bibinfo
  {title} {Three-dimensional quantum {H}all effect and metal-insulator
  transition in $\text{ZrTe}_5$},\ }\href
  {https://doi.org/10.1038/s41586-019-1180-9} {\bibfield  {journal} {\bibinfo
  {journal} {Nature}\ }\textbf {\bibinfo {volume} {569}},\ \bibinfo {pages}
  {537} (\bibinfo {year} {2019})}\BibitemShut {NoStop}%
\bibitem [{\citenamefont {Liang}\ \emph {et~al.}(2018)\citenamefont {Liang},
  \citenamefont {Lin}, \citenamefont {Gibson}, \citenamefont {Kushwaha},
  \citenamefont {Liu}, \citenamefont {Wang}, \citenamefont {Xiong},
  \citenamefont {Sobota}, \citenamefont {Hashimoto}, \citenamefont {Kirchmann}
  \emph {et~al.}}]{liang2018anomalous}%
  \BibitemOpen
  \bibfield  {author} {\bibinfo {author} {\bibfnamefont {T.}~\bibnamefont
  {Liang}}, \bibinfo {author} {\bibfnamefont {J.}~\bibnamefont {Lin}}, \bibinfo
  {author} {\bibfnamefont {Q.}~\bibnamefont {Gibson}}, \bibinfo {author}
  {\bibfnamefont {S.}~\bibnamefont {Kushwaha}}, \bibinfo {author}
  {\bibfnamefont {M.}~\bibnamefont {Liu}}, \bibinfo {author} {\bibfnamefont
  {W.}~\bibnamefont {Wang}}, \bibinfo {author} {\bibfnamefont {H.}~\bibnamefont
  {Xiong}}, \bibinfo {author} {\bibfnamefont {J.~A.}\ \bibnamefont {Sobota}},
  \bibinfo {author} {\bibfnamefont {M.}~\bibnamefont {Hashimoto}}, \bibinfo
  {author} {\bibfnamefont {P.~S.}\ \bibnamefont {Kirchmann}}, \emph {et~al.},\
  }\bibfield  {title} {\bibinfo {title} {Anomalous {H}all effect in
  $\text{ZrTe}_5$},\ }\href {https://doi.org/10.1038/s41567-018-0078-z}
  {\bibfield  {journal} {\bibinfo  {journal} {Nat. Phys.}\ }\textbf {\bibinfo
  {volume} {14}},\ \bibinfo {pages} {451} (\bibinfo {year} {2018})}\BibitemShut
  {NoStop}%
\bibitem [{\citenamefont {Liu}\ \emph {et~al.}(2021)\citenamefont {Liu},
  \citenamefont {Wang}, \citenamefont {Fu}, \citenamefont {Ge}, \citenamefont
  {Li}, \citenamefont {Xi}, \citenamefont {Zhang}, \citenamefont {Yan},
  \citenamefont {Mandrus}, \citenamefont {Yan} \emph
  {et~al.}}]{liu2021induced}%
  \BibitemOpen
  \bibfield  {author} {\bibinfo {author} {\bibfnamefont {Y.}~\bibnamefont
  {Liu}}, \bibinfo {author} {\bibfnamefont {H.}~\bibnamefont {Wang}}, \bibinfo
  {author} {\bibfnamefont {H.}~\bibnamefont {Fu}}, \bibinfo {author}
  {\bibfnamefont {J.}~\bibnamefont {Ge}}, \bibinfo {author} {\bibfnamefont
  {Y.}~\bibnamefont {Li}}, \bibinfo {author} {\bibfnamefont {C.}~\bibnamefont
  {Xi}}, \bibinfo {author} {\bibfnamefont {J.}~\bibnamefont {Zhang}}, \bibinfo
  {author} {\bibfnamefont {J.}~\bibnamefont {Yan}}, \bibinfo {author}
  {\bibfnamefont {D.}~\bibnamefont {Mandrus}}, \bibinfo {author} {\bibfnamefont
  {B.}~\bibnamefont {Yan}}, \emph {et~al.},\ }\bibfield  {title} {\bibinfo
  {title} {Induced anomalous hall effect of massive dirac fermions in
  $\text{ZrTe}_5$ and $\text{HfTe}_5$ thin flakes},\ }\href
  {https://link.aps.org/doi/10.1103/PhysRevB.103.L201110} {\bibfield  {journal}
  {\bibinfo  {journal} {Phys. Rev. B}\ }\textbf {\bibinfo {volume} {103}},\
  \bibinfo {pages} {L201110} (\bibinfo {year} {2021})}\BibitemShut {NoStop}%
\bibitem [{\citenamefont {Mutch}\ \emph {et~al.}(2021)\citenamefont {Mutch},
  \citenamefont {Ma}, \citenamefont {Wang}, \citenamefont {Malinowski},
  \citenamefont {Ayres-Sims}, \citenamefont {Jiang}, \citenamefont {Liu},
  \citenamefont {Xiao}, \citenamefont {Yankowitz},\ and\ \citenamefont
  {Chu}}]{mutch2021abrupt}%
  \BibitemOpen
  \bibfield  {author} {\bibinfo {author} {\bibfnamefont {J.}~\bibnamefont
  {Mutch}}, \bibinfo {author} {\bibfnamefont {X.}~\bibnamefont {Ma}}, \bibinfo
  {author} {\bibfnamefont {C.}~\bibnamefont {Wang}}, \bibinfo {author}
  {\bibfnamefont {P.}~\bibnamefont {Malinowski}}, \bibinfo {author}
  {\bibfnamefont {J.}~\bibnamefont {Ayres-Sims}}, \bibinfo {author}
  {\bibfnamefont {Q.}~\bibnamefont {Jiang}}, \bibinfo {author} {\bibfnamefont
  {Z.}~\bibnamefont {Liu}}, \bibinfo {author} {\bibfnamefont {D.}~\bibnamefont
  {Xiao}}, \bibinfo {author} {\bibfnamefont {M.}~\bibnamefont {Yankowitz}},\
  and\ \bibinfo {author} {\bibfnamefont {J.-H.}\ \bibnamefont {Chu}},\
  }\bibfield  {title} {\bibinfo {title} {Abrupt switching of the anomalous
  {H}all effect by field-rotation in nonmagnetic $\text{ZrTe}_5$},\ }\href
  {https://arxiv.org/abs/2101.02681} {\bibfield  {journal} {\bibinfo  {journal}
  {arXiv preprint arXiv:2101.02681}\ } (\bibinfo {year} {2021})}\BibitemShut
  {NoStop}%
\bibitem [{\citenamefont {Luo}\ \emph {et~al.}(2021)\citenamefont {Luo},
  \citenamefont {Cheng}, \citenamefont {Song}, \citenamefont {Wang},
  \citenamefont {Vaswani}, \citenamefont {Lozano}, \citenamefont {Gu},
  \citenamefont {Huang}, \citenamefont {Kim}, \citenamefont {Liu} \emph
  {et~al.}}]{luo2021light}%
  \BibitemOpen
  \bibfield  {author} {\bibinfo {author} {\bibfnamefont {L.}~\bibnamefont
  {Luo}}, \bibinfo {author} {\bibfnamefont {D.}~\bibnamefont {Cheng}}, \bibinfo
  {author} {\bibfnamefont {B.}~\bibnamefont {Song}}, \bibinfo {author}
  {\bibfnamefont {L.-L.}\ \bibnamefont {Wang}}, \bibinfo {author}
  {\bibfnamefont {C.}~\bibnamefont {Vaswani}}, \bibinfo {author} {\bibfnamefont
  {P.~M.}\ \bibnamefont {Lozano}}, \bibinfo {author} {\bibfnamefont
  {G.}~\bibnamefont {Gu}}, \bibinfo {author} {\bibfnamefont {C.}~\bibnamefont
  {Huang}}, \bibinfo {author} {\bibfnamefont {R.~H.}\ \bibnamefont {Kim}},
  \bibinfo {author} {\bibfnamefont {Z.}~\bibnamefont {Liu}}, \emph {et~al.},\
  }\bibfield  {title} {\bibinfo {title} {A light-induced phononic symmetry
  switch and giant dissipationless topological photocurrent in
  $\text{ZrTe}_5$},\ }\href {https://doi.org/10.1038/s41563-020-00882-4}
  {\bibfield  {journal} {\bibinfo  {journal} {Nat. Mater.}\ }\textbf {\bibinfo
  {volume} {20}},\ \bibinfo {pages} {329} (\bibinfo {year} {2021})}\BibitemShut
  {NoStop}%
\bibitem [{\citenamefont {Nagaosa}\ \emph {et~al.}(2010)\citenamefont
  {Nagaosa}, \citenamefont {Sinova}, \citenamefont {Onoda}, \citenamefont
  {MacDonald},\ and\ \citenamefont {Ong}}]{nagaosa2010}%
  \BibitemOpen
  \bibfield  {author} {\bibinfo {author} {\bibfnamefont {N.}~\bibnamefont
  {Nagaosa}}, \bibinfo {author} {\bibfnamefont {J.}~\bibnamefont {Sinova}},
  \bibinfo {author} {\bibfnamefont {S.}~\bibnamefont {Onoda}}, \bibinfo
  {author} {\bibfnamefont {A.~H.}\ \bibnamefont {MacDonald}},\ and\ \bibinfo
  {author} {\bibfnamefont {N.~P.}\ \bibnamefont {Ong}},\ }\bibfield  {title}
  {\bibinfo {title} {Anomalous {H}all effect},\ }\href
  {https://doi.org/10.1103/RevModPhys.82.1539} {\bibfield  {journal} {\bibinfo
  {journal} {Rev. Mod. Phys.}\ }\textbf {\bibinfo {volume} {82}},\ \bibinfo
  {pages} {1539} (\bibinfo {year} {2010})}\BibitemShut {NoStop}%
\bibitem [{\citenamefont {Karplus}\ and\ \citenamefont
  {Luttinger}(1954)}]{karplus1954}%
  \BibitemOpen
  \bibfield  {author} {\bibinfo {author} {\bibfnamefont {R.}~\bibnamefont
  {Karplus}}\ and\ \bibinfo {author} {\bibfnamefont {J.~M.}\ \bibnamefont
  {Luttinger}},\ }\bibfield  {title} {\bibinfo {title} {Hall effect in
  ferromagnetics},\ }\href {https://doi.org/10.1103/PhysRev.95.1154} {\bibfield
   {journal} {\bibinfo  {journal} {Phys. Rev.}\ }\textbf {\bibinfo {volume}
  {95}},\ \bibinfo {pages} {1154} (\bibinfo {year} {1954})}\BibitemShut
  {NoStop}%
\bibitem [{\citenamefont {Liu}\ \emph {et~al.}(2016)\citenamefont {Liu},
  \citenamefont {Yuan}, \citenamefont {Zhang}, \citenamefont {Jin},
  \citenamefont {Narayan}, \citenamefont {Luo}, \citenamefont {Chen},
  \citenamefont {Yang}, \citenamefont {Zou}, \citenamefont {Wu}, \citenamefont
  {Sanvito}, \citenamefont {Xia}, \citenamefont {Li}, \citenamefont {Wang},\
  and\ \citenamefont {Xiu}}]{liu2016zeeman}%
  \BibitemOpen
  \bibfield  {author} {\bibinfo {author} {\bibfnamefont {Y.}~\bibnamefont
  {Liu}}, \bibinfo {author} {\bibfnamefont {X.}~\bibnamefont {Yuan}}, \bibinfo
  {author} {\bibfnamefont {C.}~\bibnamefont {Zhang}}, \bibinfo {author}
  {\bibfnamefont {Z.}~\bibnamefont {Jin}}, \bibinfo {author} {\bibfnamefont
  {A.}~\bibnamefont {Narayan}}, \bibinfo {author} {\bibfnamefont
  {C.}~\bibnamefont {Luo}}, \bibinfo {author} {\bibfnamefont {Z.}~\bibnamefont
  {Chen}}, \bibinfo {author} {\bibfnamefont {L.}~\bibnamefont {Yang}}, \bibinfo
  {author} {\bibfnamefont {J.}~\bibnamefont {Zou}}, \bibinfo {author}
  {\bibfnamefont {X.}~\bibnamefont {Wu}}, \bibinfo {author} {\bibfnamefont
  {S.}~\bibnamefont {Sanvito}}, \bibinfo {author} {\bibfnamefont
  {Z.}~\bibnamefont {Xia}}, \bibinfo {author} {\bibfnamefont {L.}~\bibnamefont
  {Li}}, \bibinfo {author} {\bibfnamefont {Z.}~\bibnamefont {Wang}},\ and\
  \bibinfo {author} {\bibfnamefont {F.}~\bibnamefont {Xiu}},\ }\bibfield
  {title} {\bibinfo {title} {Zeeman splitting and dynamical mass generation in
  {D}irac semimetal $\text{ZrTe}_5$},\ }\href
  {https://doi.org/10.1038/ncomms12516} {\bibfield  {journal} {\bibinfo
  {journal} {Nat. Commun.}\ }\textbf {\bibinfo {volume} {7}},\ \bibinfo {pages}
  {12516} (\bibinfo {year} {2016})}\BibitemShut {NoStop}%
\bibitem [{\citenamefont {Choi}\ \emph {et~al.}(2020)\citenamefont {Choi},
  \citenamefont {Villanova},\ and\ \citenamefont {Park}}]{choi2020zeeman}%
  \BibitemOpen
  \bibfield  {author} {\bibinfo {author} {\bibfnamefont {Y.}~\bibnamefont
  {Choi}}, \bibinfo {author} {\bibfnamefont {J.~W.}\ \bibnamefont
  {Villanova}},\ and\ \bibinfo {author} {\bibfnamefont {K.}~\bibnamefont
  {Park}},\ }\bibfield  {title} {\bibinfo {title} {Zeeman-splitting-induced
  topological nodal structure and anomalous {H}all conductivity in
  $\text{ZrTe}_5$},\ }\href
  {https://link.aps.org/doi/10.1103/PhysRevB.101.035105} {\bibfield  {journal}
  {\bibinfo  {journal} {Phys. Rev. B}\ }\textbf {\bibinfo {volume} {101}},\
  \bibinfo {pages} {035105} (\bibinfo {year} {2020})}\BibitemShut {NoStop}%
\bibitem [{\citenamefont {Sun}\ \emph {et~al.}(2020)\citenamefont {Sun},
  \citenamefont {Cao}, \citenamefont {Cui}, \citenamefont {Zhu}, \citenamefont
  {Ma}, \citenamefont {Wang}, \citenamefont {Zhuo}, \citenamefont {Cheng},
  \citenamefont {Wang}, \citenamefont {Wan} \emph {et~al.}}]{sun2020large}%
  \BibitemOpen
  \bibfield  {author} {\bibinfo {author} {\bibfnamefont {Z.}~\bibnamefont
  {Sun}}, \bibinfo {author} {\bibfnamefont {Z.}~\bibnamefont {Cao}}, \bibinfo
  {author} {\bibfnamefont {J.}~\bibnamefont {Cui}}, \bibinfo {author}
  {\bibfnamefont {C.}~\bibnamefont {Zhu}}, \bibinfo {author} {\bibfnamefont
  {D.}~\bibnamefont {Ma}}, \bibinfo {author} {\bibfnamefont {H.}~\bibnamefont
  {Wang}}, \bibinfo {author} {\bibfnamefont {W.}~\bibnamefont {Zhuo}}, \bibinfo
  {author} {\bibfnamefont {Z.}~\bibnamefont {Cheng}}, \bibinfo {author}
  {\bibfnamefont {Z.}~\bibnamefont {Wang}}, \bibinfo {author} {\bibfnamefont
  {X.}~\bibnamefont {Wan}}, \emph {et~al.},\ }\bibfield  {title} {\bibinfo
  {title} {Large zeeman splitting induced anomalous {H}all effect in
  $\text{ZrTe}_5$},\ }\href {https://doi.org/10.1038/s41535-020-0239-z}
  {\bibfield  {journal} {\bibinfo  {journal} {npj Quantum Mater.}\ }\textbf
  {\bibinfo {volume} {5}},\ \bibinfo {pages} {1} (\bibinfo {year}
  {2020})}\BibitemShut {NoStop}%
\bibitem [{\citenamefont {Chi}\ \emph {et~al.}(2017)\citenamefont {Chi},
  \citenamefont {Zhang}, \citenamefont {Gu}, \citenamefont {Kharzeev},
  \citenamefont {Dai},\ and\ \citenamefont {Li}}]{chi2017lifshitz}%
  \BibitemOpen
  \bibfield  {author} {\bibinfo {author} {\bibfnamefont {H.}~\bibnamefont
  {Chi}}, \bibinfo {author} {\bibfnamefont {C.}~\bibnamefont {Zhang}}, \bibinfo
  {author} {\bibfnamefont {G.}~\bibnamefont {Gu}}, \bibinfo {author}
  {\bibfnamefont {D.~E.}\ \bibnamefont {Kharzeev}}, \bibinfo {author}
  {\bibfnamefont {X.}~\bibnamefont {Dai}},\ and\ \bibinfo {author}
  {\bibfnamefont {Q.}~\bibnamefont {Li}},\ }\bibfield  {title} {\bibinfo
  {title} {Lifshitz transition mediated electronic transport anomaly in bulk
  $\text{ZrTe}_5$},\ }\href {https://doi.org/10.1088/1367-2630/aa55a3}
  {\bibfield  {journal} {\bibinfo  {journal} {New J. Phys.}\ }\textbf {\bibinfo
  {volume} {19}} (\bibinfo {year} {2017})}\BibitemShut {NoStop}%
\bibitem [{\citenamefont {Fu}\ \emph {et~al.}(2020)\citenamefont {Fu},
  \citenamefont {Wang},\ and\ \citenamefont {Shen}}]{fu2020dirac}%
  \BibitemOpen
  \bibfield  {author} {\bibinfo {author} {\bibfnamefont {B.}~\bibnamefont
  {Fu}}, \bibinfo {author} {\bibfnamefont {H.-W.}\ \bibnamefont {Wang}},\ and\
  \bibinfo {author} {\bibfnamefont {S.-Q.}\ \bibnamefont {Shen}},\ }\bibfield
  {title} {\bibinfo {title} {Dirac polarons and resistivity anomaly in
  $\text{ZrTe}_5$ and $\text{HfTe}_5$},\ }\href
  {https://link.aps.org/doi/10.1103/PhysRevLett.125.256601} {\bibfield
  {journal} {\bibinfo  {journal} {Phys. Rev. Lett.}\ }\textbf {\bibinfo
  {volume} {125}} (\bibinfo {year} {2020})}\BibitemShut {NoStop}%
\bibitem [{\citenamefont {Xu}\ \emph {et~al.}(2018)\citenamefont {Xu},
  \citenamefont {Zhao}, \citenamefont {Marsik}, \citenamefont {Sheveleva},
  \citenamefont {Lyzwa}, \citenamefont {Dai}, \citenamefont {Chen},
  \citenamefont {Qiu},\ and\ \citenamefont {Bernhard}}]{xu2018temperature}%
  \BibitemOpen
  \bibfield  {author} {\bibinfo {author} {\bibfnamefont {B.}~\bibnamefont
  {Xu}}, \bibinfo {author} {\bibfnamefont {L.}~\bibnamefont {Zhao}}, \bibinfo
  {author} {\bibfnamefont {P.}~\bibnamefont {Marsik}}, \bibinfo {author}
  {\bibfnamefont {E.}~\bibnamefont {Sheveleva}}, \bibinfo {author}
  {\bibfnamefont {F.}~\bibnamefont {Lyzwa}}, \bibinfo {author} {\bibfnamefont
  {Y.}~\bibnamefont {Dai}}, \bibinfo {author} {\bibfnamefont {G.}~\bibnamefont
  {Chen}}, \bibinfo {author} {\bibfnamefont {X.}~\bibnamefont {Qiu}},\ and\
  \bibinfo {author} {\bibfnamefont {C.}~\bibnamefont {Bernhard}},\ }\bibfield
  {title} {\bibinfo {title} {Temperature-driven topological phase transition
  and intermediate {D}irac semimetal phase in $\text{ZrTe}_5$},\ }\href
  {https://journals.aps.org/prl/abstract/10.1103/PhysRevLett.121.187401}
  {\bibfield  {journal} {\bibinfo  {journal} {Phys. Rev. Lett.}\ }\textbf
  {\bibinfo {volume} {121}},\ \bibinfo {pages} {187401} (\bibinfo {year}
  {2018})}\BibitemShut {NoStop}%
\bibitem [{\citenamefont {Okada}\ \emph {et~al.}(1980)\citenamefont {Okada},
  \citenamefont {Sambongi},\ and\ \citenamefont {Ido}}]{okada1980giant}%
  \BibitemOpen
  \bibfield  {author} {\bibinfo {author} {\bibfnamefont {S.}~\bibnamefont
  {Okada}}, \bibinfo {author} {\bibfnamefont {T.}~\bibnamefont {Sambongi}},\
  and\ \bibinfo {author} {\bibfnamefont {M.}~\bibnamefont {Ido}},\ }\bibfield
  {title} {\bibinfo {title} {Giant resistivity anomaly in $\text{ZrTe}_5$},\
  }\href {https://doi.org/10.1143/JPSJ.49.839} {\bibfield  {journal} {\bibinfo
  {journal} {J. Phys. Soc. Jpn.}\ }\textbf {\bibinfo {volume} {49}},\ \bibinfo
  {pages} {839} (\bibinfo {year} {1980})}\BibitemShut {NoStop}%
\bibitem [{\citenamefont {Jones}\ \emph {et~al.}(1982)\citenamefont {Jones},
  \citenamefont {Fuller}, \citenamefont {Wieting},\ and\ \citenamefont
  {Levy}}]{jones1982thermoelectric}%
  \BibitemOpen
  \bibfield  {author} {\bibinfo {author} {\bibfnamefont {T.}~\bibnamefont
  {Jones}}, \bibinfo {author} {\bibfnamefont {W.}~\bibnamefont {Fuller}},
  \bibinfo {author} {\bibfnamefont {T.}~\bibnamefont {Wieting}},\ and\ \bibinfo
  {author} {\bibfnamefont {F.}~\bibnamefont {Levy}},\ }\bibfield  {title}
  {\bibinfo {title} {Thermoelectric power of $\text{HfTe}_5$ and
  $\text{ZrTe}_5$},\ }\href
  {https://doi.org/https://doi.org/10.1016/0038-1098(82)90008-4} {\bibfield
  {journal} {\bibinfo  {journal} {Solid State Commun.}\ }\textbf {\bibinfo
  {volume} {42}},\ \bibinfo {pages} {793} (\bibinfo {year} {1982})}\BibitemShut
  {NoStop}%
\bibitem [{\citenamefont {Tritt}\ \emph {et~al.}(1999)\citenamefont {Tritt},
  \citenamefont {Lowhorn}, \citenamefont {Littleton}, \citenamefont {Pope},
  \citenamefont {Feger},\ and\ \citenamefont {Kolis}}]{Tritt1999large}%
  \BibitemOpen
  \bibfield  {author} {\bibinfo {author} {\bibfnamefont {T.~M.}\ \bibnamefont
  {Tritt}}, \bibinfo {author} {\bibfnamefont {N.~D.}\ \bibnamefont {Lowhorn}},
  \bibinfo {author} {\bibfnamefont {R.~T.}\ \bibnamefont {Littleton}}, \bibinfo
  {author} {\bibfnamefont {A.}~\bibnamefont {Pope}}, \bibinfo {author}
  {\bibfnamefont {C.~R.}\ \bibnamefont {Feger}},\ and\ \bibinfo {author}
  {\bibfnamefont {J.~W.}\ \bibnamefont {Kolis}},\ }\bibfield  {title} {\bibinfo
  {title} {Large enhancement of the resistive anomaly in the pentatelluride
  materials $\text{HfTe}_5$ and $\text{ZrTe}_5$ with applied magnetic field},\
  }\href {https://doi.org/10.1103/PhysRevB.60.7816} {\bibfield  {journal}
  {\bibinfo  {journal} {Phys. Rev. B}\ }\textbf {\bibinfo {volume} {60}},\
  \bibinfo {pages} {7816} (\bibinfo {year} {1999})}\BibitemShut {NoStop}%
\bibitem [{\citenamefont {Zhang}\ \emph {et~al.}(2017)\citenamefont {Zhang},
  \citenamefont {Wang}, \citenamefont {Yu}, \citenamefont {Liu}, \citenamefont
  {Liang}, \citenamefont {Huang}, \citenamefont {Nie}, \citenamefont {Sun},
  \citenamefont {Zhang}, \citenamefont {Shen} \emph
  {et~al.}}]{zhang2017electronic}%
  \BibitemOpen
  \bibfield  {author} {\bibinfo {author} {\bibfnamefont {Y.}~\bibnamefont
  {Zhang}}, \bibinfo {author} {\bibfnamefont {C.}~\bibnamefont {Wang}},
  \bibinfo {author} {\bibfnamefont {L.}~\bibnamefont {Yu}}, \bibinfo {author}
  {\bibfnamefont {G.}~\bibnamefont {Liu}}, \bibinfo {author} {\bibfnamefont
  {A.}~\bibnamefont {Liang}}, \bibinfo {author} {\bibfnamefont
  {J.}~\bibnamefont {Huang}}, \bibinfo {author} {\bibfnamefont
  {S.}~\bibnamefont {Nie}}, \bibinfo {author} {\bibfnamefont {X.}~\bibnamefont
  {Sun}}, \bibinfo {author} {\bibfnamefont {Y.}~\bibnamefont {Zhang}}, \bibinfo
  {author} {\bibfnamefont {B.}~\bibnamefont {Shen}}, \emph {et~al.},\
  }\bibfield  {title} {\bibinfo {title} {Electronic evidence of
  temperature-induced lifshitz transition and topological nature in
  $\text{ZrTe}_5$},\ }\href {https://doi.org/10.1038/ncomms15512} {\bibfield
  {journal} {\bibinfo  {journal} {Nat. Commun.}\ }\textbf {\bibinfo {volume}
  {8}},\ \bibinfo {pages} {1} (\bibinfo {year} {2017})}\BibitemShut {NoStop}%
\bibitem [{\citenamefont {Zheng}\ \emph {et~al.}(2016)\citenamefont {Zheng},
  \citenamefont {Lu}, \citenamefont {Zhu}, \citenamefont {Ning}, \citenamefont
  {Han}, \citenamefont {Zhang}, \citenamefont {Zhang}, \citenamefont {Xi},
  \citenamefont {Yang}, \citenamefont {Du}, \citenamefont {Yang}, \citenamefont
  {Zhang},\ and\ \citenamefont {Tian}}]{Zheng2016Transport}%
  \BibitemOpen
  \bibfield  {author} {\bibinfo {author} {\bibfnamefont {G.}~\bibnamefont
  {Zheng}}, \bibinfo {author} {\bibfnamefont {J.}~\bibnamefont {Lu}}, \bibinfo
  {author} {\bibfnamefont {X.}~\bibnamefont {Zhu}}, \bibinfo {author}
  {\bibfnamefont {W.}~\bibnamefont {Ning}}, \bibinfo {author} {\bibfnamefont
  {Y.}~\bibnamefont {Han}}, \bibinfo {author} {\bibfnamefont {H.}~\bibnamefont
  {Zhang}}, \bibinfo {author} {\bibfnamefont {J.}~\bibnamefont {Zhang}},
  \bibinfo {author} {\bibfnamefont {C.}~\bibnamefont {Xi}}, \bibinfo {author}
  {\bibfnamefont {J.}~\bibnamefont {Yang}}, \bibinfo {author} {\bibfnamefont
  {H.}~\bibnamefont {Du}}, \bibinfo {author} {\bibfnamefont {K.}~\bibnamefont
  {Yang}}, \bibinfo {author} {\bibfnamefont {Y.}~\bibnamefont {Zhang}},\ and\
  \bibinfo {author} {\bibfnamefont {M.}~\bibnamefont {Tian}},\ }\bibfield
  {title} {\bibinfo {title} {Transport evidence for the three-dimensional
  {D}irac semimetal phase in $\text{ZrTe}_5$},\ }\href
  {https://doi.org/10.1103/PhysRevB.93.115414} {\bibfield  {journal} {\bibinfo
  {journal} {Phys. Rev. B}\ }\textbf {\bibinfo {volume} {93}},\ \bibinfo
  {pages} {115414} (\bibinfo {year} {2016})}\BibitemShut {NoStop}%
\bibitem [{\citenamefont {Pariari}\ and\ \citenamefont
  {Mandal}(2017)}]{pariari2017coexistence}%
  \BibitemOpen
  \bibfield  {author} {\bibinfo {author} {\bibfnamefont {A.}~\bibnamefont
  {Pariari}}\ and\ \bibinfo {author} {\bibfnamefont {P.}~\bibnamefont
  {Mandal}},\ }\bibfield  {title} {\bibinfo {title} {Coexistence of topological
  {D}irac fermions on the surface and three-dimensional {D}irac cone state in
  the bulk of $\text{ZrTe}_5$ single crystal},\ }\href
  {https://doi.org/10.1038/srep40327} {\bibfield  {journal} {\bibinfo
  {journal} {Sci. Rep.}\ }\textbf {\bibinfo {volume} {7}},\ \bibinfo {pages}
  {1} (\bibinfo {year} {2017})}\BibitemShut {NoStop}%
\bibitem [{\citenamefont {Tian}\ \emph {et~al.}(2019)\citenamefont {Tian},
  \citenamefont {Ghassemi},\ and\ \citenamefont {Ross}}]{tian2019dirac}%
  \BibitemOpen
  \bibfield  {author} {\bibinfo {author} {\bibfnamefont {Y.}~\bibnamefont
  {Tian}}, \bibinfo {author} {\bibfnamefont {N.}~\bibnamefont {Ghassemi}},\
  and\ \bibinfo {author} {\bibfnamefont {J.~H.}\ \bibnamefont {Ross}},\
  }\bibfield  {title} {\bibinfo {title} {Dirac electron behavior and {NMR}
  evidence for topological band inversion in $\text{ZrTe}_5$},\ }\href
  {https://doi.org/10.1103/PhysRevB.100.165149} {\bibfield  {journal} {\bibinfo
   {journal} {Phys. Rev. B}\ }\textbf {\bibinfo {volume} {100}},\ \bibinfo
  {pages} {165149} (\bibinfo {year} {2019})}\BibitemShut {NoStop}%
\bibitem [{sup()}]{supplementary}%
  \BibitemOpen
  \href@noop {} {\bibinfo {title} {See supplementary information, which
  includes {R}efs. \cite{li2016chiral, chi2017lifshitz, shahi2018bipolar,
  martino2019twodimensional,tang2019three,Chen2017Spectroscopic,liu2021induced,mutch2021abrupt,chen2015magnetoinfrared,monteiro2015magnetotransport}}}\BibitemShut
  {NoStop}%
\bibitem [{Note1()}]{Note1}%
  \BibitemOpen
  \bibinfo {note} {Here, it is assumed that $\rho _{xx}\sim \rho _{yy}$, as
  shown in a number of recent works \cite {liu2021induced,
  mutch2021abrupt}}\BibitemShut {NoStop}%
\bibitem [{\citenamefont {Pippard}(1989)}]{pippard1989magnetoresistance}%
  \BibitemOpen
  \bibfield  {author} {\bibinfo {author} {\bibfnamefont {A.}~\bibnamefont
  {Pippard}},\ }\href {https://books.google.com/books?id=D5XHMARd2ocC} {\emph
  {\bibinfo {title} {Magnetoresistance in Metals}}},\ Cambridge Studies in Low
  Temperature Physics\ (\bibinfo  {publisher} {Cambridge University Press},\
  \bibinfo {year} {1989})\BibitemShut {NoStop}%
\bibitem [{\citenamefont {Chen}\ \emph {et~al.}(2015)\citenamefont {Chen},
  \citenamefont {Chen}, \citenamefont {Song}, \citenamefont {Schneeloch},
  \citenamefont {Gu}, \citenamefont {Wang},\ and\ \citenamefont
  {Wang}}]{chen2015magnetoinfrared}%
  \BibitemOpen
  \bibfield  {author} {\bibinfo {author} {\bibfnamefont {R.}~\bibnamefont
  {Chen}}, \bibinfo {author} {\bibfnamefont {Z.}~\bibnamefont {Chen}}, \bibinfo
  {author} {\bibfnamefont {X.-Y.}\ \bibnamefont {Song}}, \bibinfo {author}
  {\bibfnamefont {J.}~\bibnamefont {Schneeloch}}, \bibinfo {author}
  {\bibfnamefont {G.}~\bibnamefont {Gu}}, \bibinfo {author} {\bibfnamefont
  {F.}~\bibnamefont {Wang}},\ and\ \bibinfo {author} {\bibfnamefont
  {N.}~\bibnamefont {Wang}},\ }\bibfield  {title} {\bibinfo {title}
  {Magnetoinfrared spectroscopy of {L}andau levels and {Z}eeman splitting of
  three-dimensional massless {D}irac fermions in $\text{ZrTe}_5$},\ }\href
  {https://link.aps.org/doi/10.1103/PhysRevLett.115.176404} {\bibfield
  {journal} {\bibinfo  {journal} {Phys. Rev. Lett.}\ }\textbf {\bibinfo
  {volume} {115}},\ \bibinfo {pages} {176404} (\bibinfo {year}
  {2015})}\BibitemShut {NoStop}%
\bibitem [{\citenamefont {Shahi}\ \emph {et~al.}(2018)\citenamefont {Shahi},
  \citenamefont {Singh}, \citenamefont {Sun}, \citenamefont {Zhao},
  \citenamefont {Chen}, \citenamefont {Lv}, \citenamefont {Li}, \citenamefont
  {Yan}, \citenamefont {Mandrus},\ and\ \citenamefont
  {Cheng}}]{shahi2018bipolar}%
  \BibitemOpen
  \bibfield  {author} {\bibinfo {author} {\bibfnamefont {P.}~\bibnamefont
  {Shahi}}, \bibinfo {author} {\bibfnamefont {D.~J.}\ \bibnamefont {Singh}},
  \bibinfo {author} {\bibfnamefont {J.~P.}\ \bibnamefont {Sun}}, \bibinfo
  {author} {\bibfnamefont {L.~X.}\ \bibnamefont {Zhao}}, \bibinfo {author}
  {\bibfnamefont {G.~F.}\ \bibnamefont {Chen}}, \bibinfo {author}
  {\bibfnamefont {Y.~Y.}\ \bibnamefont {Lv}}, \bibinfo {author} {\bibfnamefont
  {J.}~\bibnamefont {Li}}, \bibinfo {author} {\bibfnamefont {J.-Q.}\
  \bibnamefont {Yan}}, \bibinfo {author} {\bibfnamefont {D.~G.}\ \bibnamefont
  {Mandrus}},\ and\ \bibinfo {author} {\bibfnamefont {J.-G.}\ \bibnamefont
  {Cheng}},\ }\bibfield  {title} {\bibinfo {title} {Bipolar conduction as the
  possible origin of the electronic transition in pentatellurides: Metallic vs
  semiconducting behavior},\ }\href {https://doi.org/10.1103/PhysRevX.8.021055}
  {\bibfield  {journal} {\bibinfo  {journal} {Phys. Rev. X}\ }\textbf {\bibinfo
  {volume} {8}},\ \bibinfo {pages} {021055} (\bibinfo {year}
  {2018})}\BibitemShut {NoStop}%
\bibitem [{\citenamefont {Martino}\ \emph {et~al.}(2019)\citenamefont
  {Martino}, \citenamefont {Crassee}, \citenamefont {Eguchi}, \citenamefont
  {Santos-Cottin}, \citenamefont {Zhong}, \citenamefont {Gu}, \citenamefont
  {Berger}, \citenamefont {Rukelj}, \citenamefont {Orlita}, \citenamefont
  {Homes},\ and\ \citenamefont {Akrap}}]{martino2019twodimensional}%
  \BibitemOpen
  \bibfield  {author} {\bibinfo {author} {\bibfnamefont {E.}~\bibnamefont
  {Martino}}, \bibinfo {author} {\bibfnamefont {I.}~\bibnamefont {Crassee}},
  \bibinfo {author} {\bibfnamefont {G.}~\bibnamefont {Eguchi}}, \bibinfo
  {author} {\bibfnamefont {D.}~\bibnamefont {Santos-Cottin}}, \bibinfo {author}
  {\bibfnamefont {R.~D.}\ \bibnamefont {Zhong}}, \bibinfo {author}
  {\bibfnamefont {G.~D.}\ \bibnamefont {Gu}}, \bibinfo {author} {\bibfnamefont
  {H.}~\bibnamefont {Berger}}, \bibinfo {author} {\bibfnamefont
  {Z.}~\bibnamefont {Rukelj}}, \bibinfo {author} {\bibfnamefont
  {M.}~\bibnamefont {Orlita}}, \bibinfo {author} {\bibfnamefont {C.~C.}\
  \bibnamefont {Homes}},\ and\ \bibinfo {author} {\bibfnamefont
  {A.}~\bibnamefont {Akrap}},\ }\bibfield  {title} {\bibinfo {title}
  {Two-dimensional conical dispersion in ${\mathrm{zrte}}_{5}$ evidenced by
  optical spectroscopy},\ }\href
  {https://doi.org/10.1103/PhysRevLett.122.217402} {\bibfield  {journal}
  {\bibinfo  {journal} {Phys. Rev. Lett.}\ }\textbf {\bibinfo {volume} {122}},\
  \bibinfo {pages} {217402} (\bibinfo {year} {2019})}\BibitemShut {NoStop}%
\bibitem [{\citenamefont {Monteiro}\ \emph {et~al.}(2015)\citenamefont
  {Monteiro}, \citenamefont {Abanov},\ and\ \citenamefont
  {Kharzeev}}]{monteiro2015magnetotransport}%
  \BibitemOpen
  \bibfield  {author} {\bibinfo {author} {\bibfnamefont {G.~M.}\ \bibnamefont
  {Monteiro}}, \bibinfo {author} {\bibfnamefont {A.~G.}\ \bibnamefont
  {Abanov}},\ and\ \bibinfo {author} {\bibfnamefont {D.~E.}\ \bibnamefont
  {Kharzeev}},\ }\bibfield  {title} {\bibinfo {title} {Magnetotransport in
  {D}irac metals: Chiral magnetic effect and quantum oscillations},\
  }\href@noop {} {\bibfield  {journal} {\bibinfo  {journal} {Phys. Rev. B}\
  }\textbf {\bibinfo {volume} {92}},\ \bibinfo {pages} {165109} (\bibinfo
  {year} {2015})}\BibitemShut {NoStop}%
\end{thebibliography}%

\end{document}